# Anomalous Variability of Dyson Megastructures

Z. Osmanov and V. I. Berezhiani

School of Physics, Free University of Tbilisi, 0183, Tbilisi, Georgia

In the framework of the approach of Dyson megastructures, by assuming that a super-advanced civilization exists and is capable of constructing a ring-like megastructure around their host star, we have considered the observational signatures of cold (300 K) and hot (4000) K astro-engineering to answer the question: are the modern facilities capable to detect theoretically predicted fingerprints? By implying the spectral resolving power and the radial velocity methods it has been shown that the oscillation of the rings might be detected as anomalous variability of the megastructures.

## 1 Introduction

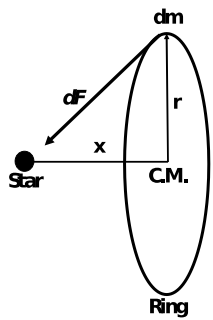

Fig. 1: The out-of-plane position of a star.

An original approach to SETI (Search for Extraterrestrial Intelligence) has been proposed by Freeman Dyson [1][1], where the author, assuming that if an extraterrestrial super advanced civilization is capable of building huge cosmic megastructures, they in principle might cloak their host star in a sphere (Dyson sphere - DS) to collect and consume the whole energy of the star. As a result, if located in the habitable zone (HZ), the DS will be visible in infrared. According to the Kardashev classification [2], the mentioned very high technology alien society is a Level-II civilization, which utilizes the whole emitted energy.

A relatively recent discovery by the Kepler mission has revived the idea of Dyson. In particular, it has been found that the object KIC8462852 (Tabby's star) revealed anomalously high dips in flux, about 20% [3], initially questioning even the possibility of shading by star dust. This mysterious object has been studied not only in optics [4] but in radio band as well [5] and in the recent study

---

[1] One should note that in 1937 the same idea has been proposed in the novel "Star Maker" by the well-known UK author and philosopher Olaf Stapledon.

in [6] it has been shown that the detected dips might originate by exocomets.

In general there have been several attempts to identify candidates of DSs on the sky [7-9], but nothing has been found. More promising results have been obtained in [10], where, based on the data of The Infrared Astronomical Satellite (IRAS), 16 objects have been observed, which potentially could be DS candidates. In this context it is worth noting that recently Zackrisson et al. (2018) [11] have considered the SETI problem in connection with the Gaia mission. In particular, they have identified the object TYC 6111-1162-1 as one of the most intriguing candidates for further study.

The search for astroengineering of Level-III civilization (capable of consuming almost the whole energy of its host galaxy) has been conducted in [12], where data of the Wide-field Infrared Survey Explorer (WISE) has been used and it was shown that among galaxies approximately 100 might be interesting candidates but further study is necessary. The similar analysis have been performed in [13] where the observational signatures of galaxies partially cloaked in DSs has been studied. The author discusses the dependence of observational fingerprints on several astrophysical parameters: ages, metallicities and initial mass functions.

It is worth noting that the search for astroengineering is only one particular direction of the Search for Extraterrestrial Intelligence (SETI), but during several decades no significant results have been obtained. This fact triggered a series of works [14-17] where the failure of conventional SETI is considered and possible solutions to the problem are actively discussed. In particular, Cirkovic & Bradbury in [17] emphasize that Dyson's approach should be re-examined and widened.

A relatively novel view to the Dysonian approach to SETI has been presented [18], where it has been assumed that Level-II civilization can colonise the nearby area of a pulsar to receive and utilise its whole emitted energy. As it has been realised, instead of a sphere the super-advanced civilization must use a ring-like structure, which if located in the HZ will be seen in the sky as infrared source. In the following work [19] the possible detection of these megastructures by modern technology has been considered and it was shown that in the neighbourhood of the solar system approximately 64±21 pulsars should be monitored. On the possible role of pulsars as beacons is discussed in [20].

In the standard Dysonian approach the megastructures are visible in the infrared spectral band, but we assume that for Level-II civilization it will be quite plausible to build a giant spherical shell inside the star's HZ (Osmanov & Berezhiani 2018 [21], henceforth OB). We have shown that for high melting temperature meta materials (in particular, for Graphene) the DS might also be detected in the optical spectrum and the size of the construction will be much smaller than the radius of DSs located in the HZ. It is worth noting that the concept of hotter DSs has been considered in [22] in the context of extremely high performance computing system. In general, the problem of detection of life is very complex and one has to take into account that there is a non-trivial connection of probability of the existence of evolved extraterrestrial life-forms and the number of biosignatures [23].

In the present manuscript we use the method developed in OB and estimate a class of stars which if surrounded by Dyson ring-like structures, oscillating by means of the gravitational interaction, might be detected by modern facilities working in the near infrared and optical spectral band. It is worth noting that complete DSs will not exhibit the oscillation character of dynamics due to the spherical symmetry of black body radiation.

The paper is organized in the following way: in Section 2, after introducing the main theoretical approach, we define observational signatures of variability of megastructures to be potentially detectable by modern technology and in Section 3 we outline the summary of our results and further possibilities how to extend the present research.

## 2 Theory and applications

In this section we partially outline the methods developed in OB[2], and by taking into account the characteristics of modern facilities, we fix the observational features of ring-like megastructures.

### 2.1 Main approach

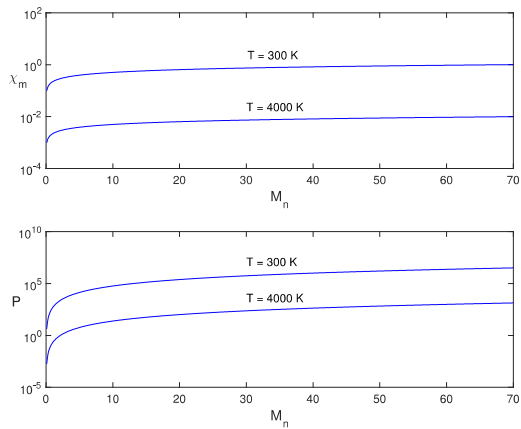

Fig. 2: Here we present the dependence of minimum value of the dimensionless amplitude (upper panel) and period of oscillations (bottom panel) as functions of central object's mass for cold and hot Dyson megastructures respectively. The set of parameters is: $\cos\theta \simeq 1$, $\alpha_1 = 1.03$, $\alpha_2 = 3.42$ and $R_p$=25000 for cold rings ($T$=300K) and $R_p$=190000 for hot rings ($T$=4000K).

---

[2] We are grateful to Prof. J. Wright for making aware of an incorrect conclusion concerning complete DSs. In particular, the fully complete DS will not lead to variability, but rings will be characterised by spectral variability.

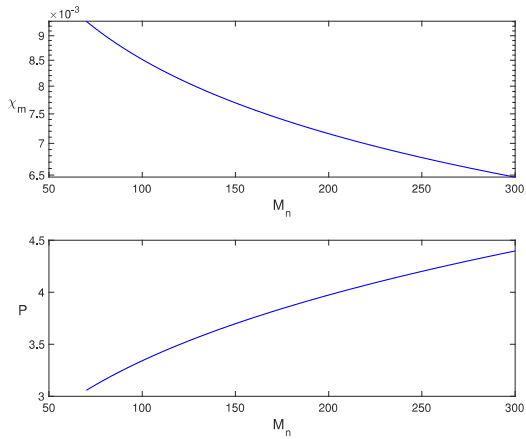

Fig. 3: Here we plot $\chi_m$ (upper panel) and $P$ (bottom panel) with $M_n$ for hot megastructures. The set of parameters is the same as in Fig. 2, except $\alpha_1 = 2.3 \times 10^4$ and $\alpha_2 = 1$.

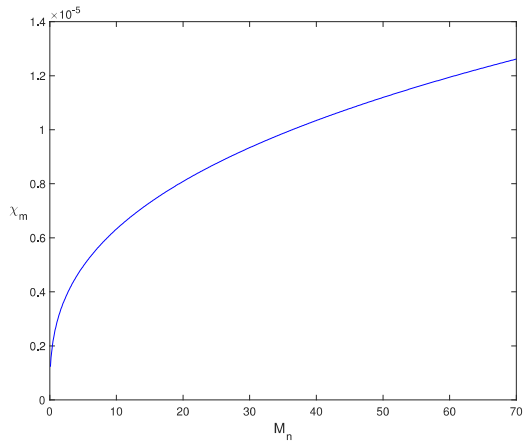

Fig. 4: The behaviour of versus the stellar mass is shown. The set of parameters is $\upsilon_m = 1$m/sec, $\alpha_1 = 1.03$, $\alpha_2 = 3.42$ and $T = 4000$K.

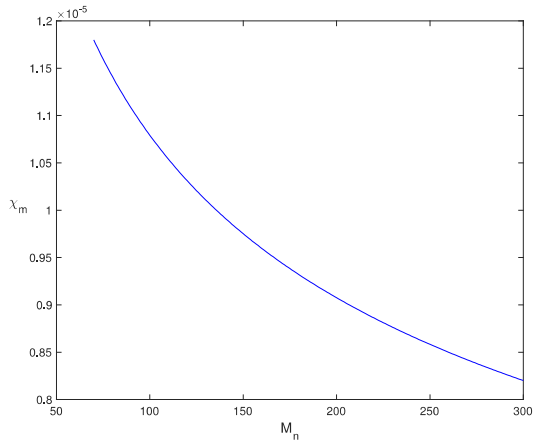

Fig 5: Here we plot the dependence $\chi_m(M_n)$. The set of parameters is the same as in the previous

figure, except $\alpha_1=2.3\times 10^4$, $\alpha_2=1$,

McInnes (see [24]) has considered the stability problem of a thin ring surrounding a central point mass. It has been shown that the in-plane motion undergoes unstable dynamics and to restore the equilibrium position (ring's centre coincident with the central mass) requires power much less than the star's luminosity [18]. On the other hand, the out-of-plane motion of the ring becomes stable only by means of the gravitational interaction of the ring and the central object. In Fig. 1 we demonstrate the out-of-plane position of the system with the corresponding displacement $x$ between the star and the C.M. (the centre of mass of the ring). One can straightforwardly show that for small amplitude oscillations the restoring force

$$F_x \simeq -G\frac{Mm}{R^2}x, \tag{1}$$

leads to the following equation of motion

$$\frac{d^2\xi}{dt^2} + \omega^2 \xi = 0, \tag{2}$$

where $G$ is the gravitational constant, $M$ and $m$ are the star's and ring's masses respectively, $R$ represents ring's radius, $\xi \equiv x/R$ and $\omega = \sqrt{GM_{st}/R^3}$ is the oscillation frequency. It is worth noting that circumstellar rings are more stable than shells.

By assuming the black body radiation of the ring and taking into account that it will emit from the outer as well as the inner surfaces, the radius of the megastructure is given by

$$R = \left(\frac{L}{8\pi\sigma T^4}\right)^{1/2} \simeq$$

$$\simeq 1.22 \times \left(\frac{L}{L_\odot}\right)^{1/2} \times \left(\frac{300K}{T}\right)^2 AU, \tag{3}$$

where $L$ is the luminosity of the star, $\sigma \approx 5.67 \times 10^{-5}$ erg/(cm$^2$ K$^4$) denotes the Stefan-Boltzmann's constant and $L_\odot \approx 3.83\times 10^{33}$ ergs s$^{-1}$ is the solar luminosity. In OB we have considered the possibility of the so called hot DSs. In particular, it has been discussed that if our technological civilization, which is supposed to be almost Level-I, can produce Graphene with very high melting temperature 4510K [25], there might not be a serious problem to perform the same job in super advanced societies of Level-II. From the aforementioned expression it is clear that the hot megastructure with $T$=4000K will have radii of by two orders of magnitude smaller than the lengthscales of cold megastructures located in the HZ. In OB we have also considered how plausible a cooling system might be to maintain habitable conditions inside the shell. By examining the typical values of coefficient of performance (COP) of modern cooling devices, it has been shown that an engine, to compensate the heat flux to the cold area, should process the power (normalised to the Solar power) of the order of P $\simeq 2.5 \times 10^5 \xi \frac{\Delta T/h}{370\ K\ cm^{-1}}$, where we have taken into account the temperature difference, $\Delta T$=(4000-300)K=3700K between the outer surface and an internal surface bordering inhabitants, $h$=10cm is the corresponding effective thickness and

$\xi \simeq 1$ is a dimensionless factor depending on geometry of free space inside the shell. For more details please see the corresponding discussion in OB[3]. As it is clear from the value of $P_e$, there is a tiny fraction of energy received from the host star, which means that the process of cooling is quite realistic. From Eq. (2) one can show that the ring will oscillate with the following characteristic period

$$P = 2\pi\sqrt{\frac{R^3}{GM}} \simeq 490 \times \left(\frac{M_\odot}{M}\right)^{1/2} \times$$

$$\times \left(\frac{L}{L_\odot}\right)^{3/4} \times \left(\frac{300K}{T}\right)^3 days,$$

(4)

where $M_\odot \simeq 2 \times 10^{33}$g is the Solar mass. As it is evident from this expression, for hot rings $T \simeq 4000$K, the oscillation period becomes anomalously short (~5hours) which might be a good sign to distinguish the artificial constructions from astrophysical objects by studying the variability of spectral features.

## 2.2 Variability of megastructures

In this subsection we consider variability of emission pattern caused by out-of-plane oscillation of ring-like megastructures.

It is obvious that due to the relativistic Doppler effect the spectral pattern of the megastructure will be variable. In particular, the observed wavelength of the megastructure is given by [26]

$$\lambda = \lambda_s \frac{1 + v\cos\theta/c}{\sqrt{1 - v^2/c^2}},$$

(5)

where $\lambda_s$ is the wavelength of the waves the ring emits, v is the its velocity and $\theta$ is an angle of velocity direction measured in the frame of the observer. After taking into account the fact that the motion is non-relativistic, from Eq. (5) one can straightforwardly derive the maximum wavelength difference

$$\Delta\lambda \simeq \lambda_s \frac{2v_m \cos\theta}{c},$$

(6)

$v_m = \omega A$ is the velocity amplitude, $\omega = 2\pi/P$ denotes the frequency, $A = \chi R$ is the spatial amplitude of oscillations and $\chi \ll 1$ is a dimensionless parameter.

Spectrographs of modern high sensitivity telescopes are characterised by high values of the resolving power, Rp ≡ λ/Δλ. For example, the European Southern Observatory's (ESO) Very Large Telescope (VLT) has the ultra high spectral resolving power up to 25000 for cold rings (*T*=300K, λ ≈ 9.6µm) and up to 190000 for hot rings (*T*=4000K, λ ≈ 725nm)[4]. Generally speaking, one has to note that there is a nontrivial mass-luminosity relation in main sequence stars. For example, luminosities of stars for a wide range of masses M > 0.1M⊙ up to ~ $10^2$M⊙ up to can be approximated as [27]

$$L \simeq \alpha_1 \times L_\odot \times \left(\frac{M}{M_\odot}\right)^{\alpha_2},$$

(7)

with $\alpha_1 \simeq 1.03$ and $\alpha_2 \simeq 3.42$. On the other hand, massive stars have tendency to emit in the extreme regime, when the luminosity is of the order of the Eddington limit [26]

$$L_{Edd} \simeq 2.3 \times 10^4 \times L_\odot \times \frac{M}{M_\odot},$$

(8)

thus, for this class of $\alpha_1 = 2.3 \times 10^4$ and $\alpha_2 = 1$. For approximately M < 70M⊙ the luminosity-mass relation is determined by Eq. (7) and for higher mass stars - by Eq. (8).

---

[3] In OB there are misprints: $P_c$ and $P_e$ should be dimensionless and $\Delta T/h = 7$K cm$^{-1}$ instead of 70K cm$^{-1}$.

By combining the aforementioned relations with the natural condition of detectability

$$\frac{c}{2v_m \cos\theta} \leq Rp,$$

(9)

one obtains the constraint on the dimensionless spatial amplitude of oscillations

$$\chi \geq \frac{c}{2Rp\, T \cos\theta} \times \left(\frac{L}{8\pi\sigma G^2 M^2}\right)^{1/4} \simeq 0.22 \times$$

$$\times \frac{\alpha_1^{1/4}}{\cos\theta} \times \frac{25000}{Rp} \times \frac{300K}{T} \times \left(\frac{M}{M_\odot}\right)^{\frac{\alpha_2-2}{4}}. \quad (10)$$

In Fig. 2 we show the behaviour of minimum value of the dimensionless amplitude (upper panel) and period of oscillations (bottom panel) with the central object's mass for cold and hot rings respectively. The set of parameters is cosθ≈1, $\alpha_1$ =1.03, $\alpha_2$ =3.42 and Rp=25000 for cold rings (T = 300K) and Rp = 190000 for hot rings (T = 4000K). As it is clear from the plots, the search for cold megastructures' variability makes sense only for relatively small masses of stars. In particular, for just M = 10M⊙, the corresponding variability timescale is of the order of 160 yrs, but we do not have a long history of observing the sky in the infrared window and therefore the data is not enough. For masses in the 0.1M⊙ < M < 6M⊙ the search for cold rings might have sense because the period might be of the order of 50yrs. On the other hand, for this case, the values of χ are bigger than 0.1 and therefore, it seems not to be very realistic.

---

[4] www.eso.org/public/teles-instr/paranal-observatory/vlt/

Unlike the cold Dyson megastructures, the search for the variability of hot rings might be more promising. For almost the whole range of masses considered in Fig. 2 the period of oscillations is within a decade and the minimum value of oscillation amplitude potentially detectable by ESO's VLT instruments does not exceed ~ $2 \times 10^{-2}$ (for M = 70M☉). For the massive stars, with luminosities of the order of the Eddington limit, there is no need to consider cold megastructures because of the enormous values of variability timescales. In Fig. 3 we show the behaviour of $\chi_m$ and P versus the normalised mass for hot megastructures. The set of parameters is the same as in Fig. 2, except $\alpha_1 = 2.3 \times 10^4$ and $\alpha_2 = 1$. The variability period in this case be- came continuously decreasing function, which is a direct result of the fact that $\alpha_2 - 2$ is negative and the corresponding minimum values of $\chi$ are of the order of $10^{-2}$.

The aforementioned spectral method has limitation: even for the most prospective hot Dyson megastructures the minimum value of dimensionless amplitude of oscillations is of the order of $2 \times 10^{-3}$. On the other hand, if the material the ring is made of is characterised by the absorption lines, then the so-called radial velocity method can be implied quite efficiently [28]. By taking into account that sensitivity of radial velocity measurement of modern facilities is less than $1 m/s$, one can straightforwardly show that $\chi$ must satisfy the following condition to be detectable:

$$\chi \geq \frac{v_m}{T} \times \left(\frac{L}{8\pi\sigma G^2 M^2}\right)^{1/4} \simeq 1.22 \times 10^{-6} \times \alpha_1^{1/4} \times$$

$$\times \frac{v_m}{1 m/sec} \times \frac{4000K}{T} \times \left(\frac{M}{M_\odot}\right)^{\frac{\alpha_2 - 2}{4}}. \quad (11)$$

Since the period of oscillations of cold megastructures for most of the stars is extremely large, we considered only hot rings. In Fig. 4 the dependence $\chi_m(M_n)$ is presented. The set of parameters is $v_m = 1$m/sec, $\alpha_1 = 1.03$, $\alpha_2 = 1$ and T = 4000K. It is clear that this method is very efficient compared to the spectral analysis method. The corresponding minimum dimensionless amplitude lies in the interval $1.2 \times 10^{-6} - 10^{-5}$, which is by several orders of magnitude smaller than in the previous method. The similar result is obtained for massive stars (see Fig. 5), the minimum values of $\chi$ are of the order of $10^{-5}$. All these mean that the search for hot Dyson megastructures seems to be very prospective.

Generally speaking, the search for the super advanced alien astro-engineering requires focusing on the search for the anomalous behaviour of variability. In particular, for cold rings, the timescale varies from days to years (from K stars up to A type stars) up to hundreds of years and higher (B and O type stars). On the other hand, the stars of F, M, S or C class although are characterised by long scale periodicity, they have either very high temperatures (7000K for F stars) of very high luminosities (M, S and C).

Considering hot rings (T~4000K) around stars from type M to type - A, one can show that the variability timescales vary from several minutes (A) to days (M). On the other hand, it is well known that ZZ Ceti are characterised by pulsation period of the order of minutes, but these objects are extremely hot. Therefore, two variables with the same timescales (minutes) might be easily distinguished by their temperatures. Classical Cepheids have the pulsation period of the order of several days, but still, their temperature is higher than the temperature of hot rings. Longer timescale of the order of hundreds of days or even several years will be the typical periods of B and O type stars, but they are extremely luminous with very high temperatures, incompatible with the temperatures of hot megastructures.

Therefore, to summarise the aforementioned results, it is significant to notice that in order to identify interesting objects on the sky, one has to look for anomalous behaviour of variability.

## 3 Conclusion

In the framework of the paradigm that the conventional SETI approach should be extended we have assumed that the level-II civilization is capable of building a huge megastructures not only in the stars' HZs but inside them. As it has been shown, by means of the gravitational interaction the megastructure will be oscillating around the equilibrium position.

By taking into account the technical characteristics of modern facilities, we have shown that the oscillation might lead to the spectral variability of rings potentially detectable by telescopes. Two major methods have been examined: by using the maximum spectral resolving power of the VLT it has been shown that cold megastructures cannot be easily detected by variability pattern. Unlike the cold rings, the hot megastructures might be detected by spectral variability method. The second and the most important approach - the radial velocity method - has been implied and it has been argued that the search for the Dyson ring-like megastructures are very promising, but to find interesting candidates one should focus on anomalous behaviour of variability.

In general, the motivation of constructing huge megastructures might be different. In the present work we have considered that the astro-engineering is used for energy consumption and habitation, but this is not the only reason of building the megastructures. For example, it can also be used for ultra-high performance computation [22] which requires very low ambient temperature, or still for habitation, but with different biology [29] changing the HZ concept and consequently the length scales of rings.

**Acknowledgments**

The authors are grateful to Prof. J. Wright for interesting suggestions and comments. The research was supported by the Shota Rustaveli National Science Foundation grant (DI-2016-14).